\documentclass[aps,prl,preprint,groupedaddress,showpacs]{revtex4-1}
\usepackage{natbib}
\usepackage{graphicx}
\usepackage{mathtools}
\usepackage{mathrsfs}
\usepackage{dsfont}
\usepackage{bm}

\begin{document}

\title{Bloch Sphere Catastrophes}


\author{Samuel A. Eastwood}
\email[]{samuel.eastwood@monash.edu}
\affiliation{School of Physics, Monash University, Victoria 3800, Australia}

\author{David M. Paganin}
\affiliation{School of Physics, Monash University, Victoria 3800, Australia}

\author{Timothy C. Petersen}
\affiliation{School of Physics, Monash University, Victoria 3800, Australia}

\author{Michael J. Morgan}
\affiliation{School of Physics, Monash University, Victoria 3800, Australia}


\date{\today}

\begin{abstract}
Caustics are optical phenomena which occur when a family of rays creates an envelope of divergent intensity.  Here we show that caustic surfaces also appear when a real or complex field is mapped to its order parameter manifold.  We study these structures in the context of spin-$1/2$ fields, where the order parameter manifold is the Bloch sphere. These generic structures are a manifestation of catastrophe theory and are stable with respect to perturbations.  The corresponding field configurations are also stable and represent a new type of topological defect. Equations governing the conditions for their existence and unfolding  are derived. 
\end{abstract}


\maketitle

\section{Introduction}
Order parameters are pivotal to the study and classification of topological defects and phase transitions~\cite{sethna}. Many important areas of physics can be understood using this construct, e.g., crystal growth~\cite{mermin}, quantum computing ~\cite{golovach},  nematic liquid crystals~\cite{liquidcrystal}, magnetic textures, such as Skyrmions~\cite{ pflederer,muhlbauer2009skyrmion,Ezwana}, multiferroics~\cite{wang2003epitaxial, kimura2003,chu2008electric}, superconductors~\cite{PhysRevLett.110.216405, PhysRevLett.108.207004} and particle cosmology~\cite{particlecomo}, to name but a few. The order parameter may also be viewed as a function that maps points in the physical space to the order parameter space. The topological properties of the order parameter space are important in the study of defects, since they govern the stability of defects in the field~\cite{ruben2010}. 

\begin{figure}[t]
	\centering
		\includegraphics[width=8.6cm]{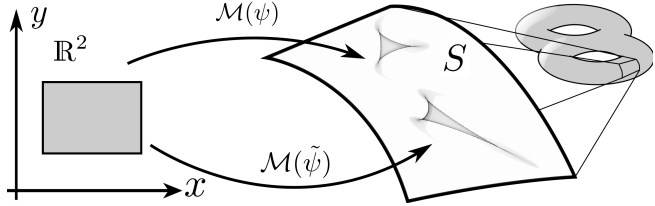}
		\caption{Schematic showing how the field $\psi$ in an $\mathds{R}^2$ patch is mapped to its order parameter manifold, leading to a 
catastrophe on $S$. Here $S$ is a local patch of the manifold, which may possess a non-trivial global topology. Due to the intrinsic 
stability of the catastrophe, a mapping of the patch of the perturbed field, $\tilde{\psi}$,  would yield a deformed catastrophe. The 
features of the field that are mapped to the catastrophe persist  in the presence of perturbations. \label{mapping}}
\end{figure}

Here we demonstrate the existence of a new type of singularity formed when a field is mapped to its order--parameter space; these defects are governed 
by the framework of catastrophe theory~\cite{arnold1975,  saunders, poston}. Thom's theorem~\cite{thom1983} tells us that catastrophes 
are stable with respect to perturbation; consequently features of the field that map to a catastrophe on the order--parameter 
space manifold must also be stable to perturbations. This singularity represents a new type of topological defect, which we refer to as an {\itshape order--parameter catastrophe defect} (see Fig.~\ref{mapping}).  

Such catastrophes possess their own topology regardless of the order parameter manifold on which they exist. Their stability is intrinsic to 
the catastrophe defect, rather than depending on the topology of the order parameter manifold.  This implies that local regions of the field, 
which map to the order parameter catastrophe defect, transcend the topology of the order parameter space, due to the intrinsic topology 
of the catastrophe itself.  Catastrophe theory may be applied to any system whose order parameter space is ``tattooed'' with such 
catastrophes. 

Catastrophes themselves are very generic structures, appearing naturally in many physical systems~\cite{hannay1982, berry1, 
berry2, nye1978, nye2003a,nye2003b,nye2006, petersen2013, simula2013, thom1977, maeda1994, arnold1975}. Hence the concepts presented here will be applicable to a broad range of phenomena. 
Although the  fundamental idea is very general,  we choose to contextualize the concept of order--parameter catastrophes by applying it 
to a 2D spin--$1/2$ field, for which the corresponding order parameter space is the Bloch sphere. 

\section{Bloch sphere mapping}
Consider the general wave--function of a two--component spinor
\begin{equation} \label{wavefunction}
\left| \Psi(\mathbf{x},t) \right\rangle = \psi_0(\mathbf{x},t) \left| 0 \right\rangle  + \psi_1(\mathbf{x},t) \left| 1 \right\rangle,
\end{equation}
where $\mathbf{x}$ is the spatial variable and $t$ the time coordinate; $\psi_0$ and $\psi_1$ are the probability amplitudes of the spin up state $\left| 0 \right\rangle $ and spin down state $\left| 1 \right\rangle $, respectively. 
Without loss of generality we may assume $|\psi_0|^2+|\psi_1|^2 = 1$, and consider only pure spin states that map to the surface of the Bloch sphere. A single spin state is represented graphically as a unit position vector of the Bloch sphere, denoted by $\left| \mathscr{B} \right\rangle$. It is parametrized by the spherical polar angles $(\phi, \theta)$, as
\begin{equation} \label{Blochvector}
			\begin{split} 
					\left| \mathscr{B} \right\rangle = e^{i\chi} [ e^{-i\phi(x,y)/2}&\mbox{cos}\left( \theta(x,y)/2 \right) \left|0\right\rangle \\ + e^{i\phi(x,y)/2}&\mbox{sin}\left(  \theta(x,y)/2 \right) \left|1\right\rangle ],
			\end{split}
\end{equation}
where  $\chi$ is the global phase of the Bloch vector~\cite{PhysRevB.67.094510, PhysRevA.72.012315, nielsen2010quantum}. The global phase factor $\exp(i \chi)$ is an unobservable of the system.

Here we are concerned with the mapping of the Bloch vector, located at every point in coordinate space, to the Bloch sphere. Consider the mapping of an $(x,y)$--patch of Bloch vectors, corresponding to a spin--$1/2$  field, to the surface of the Bloch sphere. If the patch contains a texture defect it will wrap the surface of the Bloch sphere an integer number of times~\cite{PhysRevB.79.245331,PhysRevLett.100.180403,makela2003,mermin}. If the field is perturbed locally, the texture structure will not be destroyed, since it is topologically protected. 
Instead, the mapping is altered in such a way that a local $(x,y)$--patch may be sheared, dilated or rotated by the mapping, whilst still maintaining the condition that the surface of the Bloch sphere remains wrapped. To accommodate transforms of this type, small regions on the Bloch sphere must fold back onto themselves, resulting in a fold catastrophe.
This highlights the ubiquity of order--parameter catastrophes; a continuous perturbation of a texture defect can yield a caustic ``tattoo'' on the Bloch sphere (see Fig.~\ref{mapping}).  In the general case, when a texture defect exists in a spin field, Bloch--sphere catastrophes will also be present. However, 
the converse is not necessarily true. 

\section{Conditions for a catastrophe.}
The explicit form of the Bloch sphere map is determined by the inverse of Eq.~\eqref{Blochvector}, given by
\begin{align} 
	\phi(x,y) &=   \mbox{arg}\left(\psi_1/\psi_0\right), \label{phitheta1} \\
	\theta(x,y) &=  \mbox{arccos}\left( \eta \right). \label{phitheta2}
\end{align}
The quantity $\eta = \psi_0^*\psi_0 - \psi_1^*\psi_1$ is the spin asymmetry, which represents the difference in probability density of the spin up and down components. The Bloch sphere image is a 2D histogram of $\phi$ and $\theta$, each of which are calculated for every $(x, y)$ point of the spin field. The mapping takes a patch of the $(x,y)$--plane and calculates the complex functions $\psi_0(x,y)$ and $\psi_1(x,y)$. 
Together these functions completely define the spin--$1/2$ field over the Euclidean patch in $\mathds{R}^2$ These states are then mapped to the corresponding $(\phi,\theta)$--patch on the Bloch sphere. The density of the patch on the Bloch sphere at a particular point is proportional to the number of times the spin--$1/2$ field takes on the values of that particular state. 
\begin{figure}[t]
	\centering
		\includegraphics[width = 8.6cm]{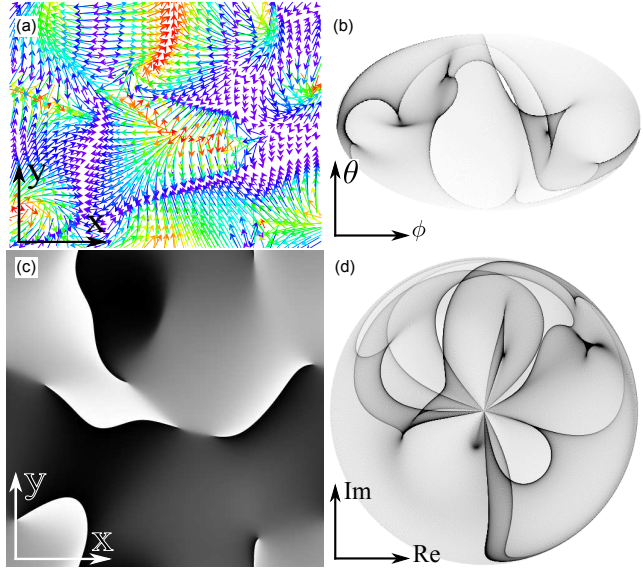}
	\caption{(a) Bloch vectors of a random spin--$1/2$ field distributed in real space with the color corresponding to their $z$--component. (b) Mollweide projection of the Bloch sphere map. Here the density of the plot is proportional to the number of spin states in the region of the spin--$1/2$ field shown in (a). (c) The Bloch sphere coordinate $\phi$ displayed as a 2D function of $(x,y)$. (d) The $\Gamma$ map, defined in the main text, of the spin field in (a). This map can be thought of as a projection mapping the Bloch sphere to the Argand plane.\label{4panel}}
\end{figure}

A randomly varying spinor wavefunction was generated by low--pass filtering an image of white noise for each spin component, producing the field shown in Fig.~\ref{4panel}(a). By applying Eqs.~\eqref{phitheta1} and~\eqref{phitheta2} the field was mapped to the Bloch sphere to produce Fig.~\ref{4panel}(b).  The presence of several phase vortices can be seen in the Bloch sphere coordinate $\phi(x,y)$, shown in real space in Fig.~\ref{4panel}(c). These screw--type topological defects come from the vortices in the phase of both components in Eq.~\eqref{wavefunction}, which is preserved when $\phi(x,y)$ is calculated. 

A Bloch sphere singularity corresponds to a many--to--one mapping. This occurs when the mapping becomes singular, hence the Jacobian determinant of the mapping must vanish. The Jacobian is defined by
\begin{equation} \label{jacobian}
J(x,y) = \begin{pmatrix} \frac{\partial\phi}{\partial x} & \frac{\partial \phi}{\partial y} \\[0.4em] \frac{\partial \theta}{\partial x}  &  \frac{\partial \theta}{\partial y}  \end{pmatrix}.
\end{equation}
Setting $\mbox{det}[J(x,y)]=0$  gives the conditions for  a particular spin state to map to a catastrophe, i.e.
\begin{equation} \label{detj}
\left[ \left(|\psi_0|^2 \mathbf{j}_2 - |\psi_1|^2 \mathbf{j}_1\right) \times \nabla \eta \right] \cdot \hat{\bm{z}} = 0,
\end{equation}
where $\mathbf{j}_1$ and $\mathbf{j}_2$ are the probability current densities of $\psi_0$ and $\psi_1$, respectively, and $\hat{\bm{z}}$ is a unit vector in the $z$ direction. It can be shown that, up to a scaling factor, Eq.~\eqref{detj} is equivalent to the $z$ component of the curl of the probability current density of a single complex scalar wave-function of the form
\begin{equation} \label{delta}
\Gamma(x,y) = \sqrt{\eta}\exp{(i\phi)}.
\end{equation}
Equation~\eqref{delta} is a projection that maps the Bloch sphere to the Argand plane; this ``$\Gamma$ map'' is shown in Fig.~\ref{4panel}(d). Zeros of $\Gamma$ are mapped to the equator of the Bloch sphere, while the entire equator is mapped to the the origin of the Argand plane. Each hemisphere is  projected over  a unit disc, with the poles mapping to the disc's circumference at $|\Gamma| = 1$. However, each hemisphere of the Bloch sphere is rotated relative to the other by $\pi/2$ radians; this is a direct result of the abrupt phase change $\Gamma$ acquires when $\eta$ changes sign.

The construction of $\Gamma$ allows us to express the equation governing order--parameter space defects of the spin--$1/2$ field in terms of the vorticity of $\Gamma$. The current density of $\Gamma$ has the form $\eta \nabla \phi$, which may be interpreted as an equation describing the flow of spin asymmetry. Catastrophes are mapped to the Bloch sphere along points where the vorticity of $\Gamma$ vanishes. 
\begin{figure}[t]
		\includegraphics[width = 8.6cm]{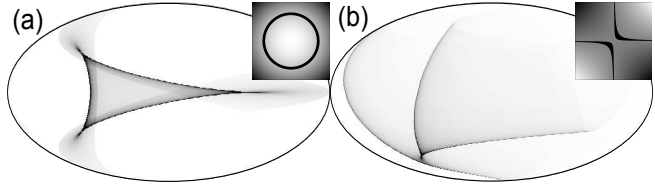}
		\caption{Mollweide plots of (a) elliptic, and (b) hyperbolic umbilic Bloch sphere catastrophes. The insets of each figure show the Jacobian determinant of the mapping, with their zeros highlighted in black. When the zeros of the Jacobian form an ellipse, the map of that region forms an elliptic umbilic catastrophe. Similarly, when the zeros form a hyperbolic curve, the field maps to the hyperbolic umbilic catastrophe.\label{zeros}}
\end{figure}

\section{Relation to the canonical form}
Catastrophes are associated with an instability of a system. Any potential function which governs the behavior of a system will consist of state variables, in this case $(x,y)$, and a set of parameters called the control variables. Degenerate critical points of a potential lead to instabilities of the system because a bifurcation may occur at these points when there is a small change in the control variables.  Catastrophes exist in the space of the control variables of the potential function. The dimensionality of the control space is determined by the number of control parameters of the potential function, i.e. the co--dimension. The {\itshape bifurcation set} defines a caustic surface embedded in control space; this corresponds to the set of values of the control variables at which bifurcations of the critical points in the potential function occur.
 
Thom's theorem describes a number of so-called {\itshape elementary catastrophes}, these being all possible topologically distinct catastrophes that occur when the co-dimension is less than or equal to four~\cite{thom1983}. 
Every elementary catastrophe has an associated function known as the canonical form~\cite{thom1983,arnold1975, saunders, poston}.  Therefore if a local patch of a spin--$1/2$ field maps to a particular elementary catastrophe on the Bloch sphere, the field itself must have a relation to the canonical form of the catastrophe.  The Bloch sphere catastrophe is embedded in the coordinates ($\phi,\theta)$ of the sphere.  However, the catastrophe bifurcation set lies in the control space. In essence, the Bloch sphere catastrophe is a slice though the control space of the bifurcation set, which has also been subject to a continuous deformation, rotation and translation. 

In catastrophe theory, a function has a degenerate critical point when the Hessian determinant vanishes. Points of the spin field where the Jacobian determinant vanishes map to a catastrophe. Therefore the Hessian of the canonical form of the catastrophe, $V(x,y)$, is locally equivalent to the Jacobian matrix of the mapping, i.e.
\begin{equation} \label{hessian}
\begin{pmatrix} \frac{\partial\tilde{\phi}}{\partial x} & \frac{\partial\tilde{\phi}}{\partial y} \\[0.4em] \frac{\partial \tilde{\theta}}{\partial x}  &  \frac{\partial \tilde{\theta}}{\partial y}  \end{pmatrix} = \begin{pmatrix} \frac{\partial^2 V}{\partial x^2} & \frac{\partial^2 V}{\partial y \partial x} \\[0.4em] \frac{\partial^2 V}{\partial x \partial y} & \frac{\partial V}{\partial y^2} \end{pmatrix},
\end{equation}
where $\tilde{\phi}$ and $\tilde{\theta}$ are the transformed coordinates of the Bloch sphere, which account for the local translation and rotation of the catastrophe from its canonical form.
To account for smooth deformations of the catastrophe of the Bloch sphere mapping, the coefficients of the transformations of $\phi$ and $\theta$  must be functions of the state space coordinates $(x,y)$. Since the transformation matrix has an inverse, this allows us to express the Bloch sphere coordinates in terms of the transformed coordinates as
\begin{equation} \label{coordtransform}
\begin{pmatrix}  \phi \\ \theta \end{pmatrix} = \begin{pmatrix} a(x,y) & b(x,y) \\ c(x,y) & d(x,y), \end{pmatrix} \begin{pmatrix} \tilde{\phi} \\ \tilde{\theta} \end{pmatrix},
\end{equation}
where the inverse transformation matrix elements $a$, $b$, $c$ and $d$ are arbitrary functions of $(x,y),$ which encapsulate the local rotations, translations or deformations of the coordinate system.

Given Eqs.~\eqref{hessian} and~\eqref{coordtransform} we find that $\phi(x,y)$ and $\theta(x,y)$ may be expressed in terms of the partial derivatives of the canonical form of the catastrophe:
	\begin{align} 
				\phi(x,y) & = a(x,y) V_x(x,y) + b(x,y) V_y(x,y) \label{gradmap1} \\
				\theta(x,y) & = c(x,y) V_x(x,y) + d(x,y) V_y(x,y)	\label{gradmap2}	
	\end{align}
where $V_x \equiv \partial V / \partial x$ and $V_y\equiv \partial V / \partial y$. Equations~\eqref{gradmap1} and~\eqref{gradmap2} show that the Bloch sphere catastrophe is indeed a form of gradient mapping of the canonical form of the catastrophe.  If a small patch on the spin field in state space were to map to a catastrophe, the field in that patch of state space must relate to the canonical form of the catastrophe. The field in these local regions has the functional form:
\begin{equation}
		\begin{aligned} \label{Blochgerm}
				\left| \Psi \right\rangle = e^{-\frac{i}{2}[(a V_x + b V_y)}\mbox{cos}&\left[(c V_x + d V_y)/2 \right] \left|0\right\rangle + \\
				e^{\frac{i}{2}(a V_x + b V_y)}\mbox{sin}&\left[(c V_x + d V_y)/2 \right] \left|1\right\rangle.
		\end{aligned}
\end{equation}

The Bloch sphere coordinates $(\phi,\theta)$ are deformed gradient maps of the canonical form of the catastrophe; the structure of the zeros of the Jacobian determinant determines the type of elementary catastrophe to which the field will map. In essence, the lines of zeros of the Jacobian determinant determine where to fold a patch in $\mathds{R}^2$, as it is mapped to the Bloch sphere. The way in which the patch is folded yields a particular type of catastrophe. 
Examples of elliptic and hyperbolic umbilic catastrophes are shown in Fig.~\ref{zeros}, along with the geometry of each catastrophe's Jacobian determinant given in the insets. The canonical form of the elliptic umbilic catastrophe is $x^3 - xy^2$,  where the null set for the Hessian determinant of this form corresponds to the equation of an ellipse, i.e. the zeros of the Jacobian determinant map to an elliptic umbilic catastrophe. Similarly the Hessian determinant of the hyperbolic umbilic catastrophe is the equation of a hyperbola, given by its canonical form of $x^3 + y^3$.  We have shown that the spin field is related to the canonical form of the catastrophe by a gradient map, therefore perturbations to the spin--$1/2$ field behave as a particular unfolding of the canonical form. 

\section{Stability and unfolding of Bloch sphere catastrophes.}
A Bloch sphere catastrophe is a particular cross--section of the full bifurcation set surface.  Catastrophes may undergo a process  known as {\itshape unfolding}. This occurs when the potential function of any given catastrophe is subject to a perturbation in such a way that the slice of the bifurcation set that is viewed on the Bloch sphere changes. In this picture we see that the topological stability of a Bloch sphere catastrophe defect arises because every catastrophe already possesses a topology that is fixed by its bifurcation set. A perturbation to the spin field  only serves to unfold the Bloch sphere catastrophe, whence we observe different slices through the bifurcation set of the catastrophe. 

\begin{figure*}[t]
	\centering
		\includegraphics[width = \textwidth]{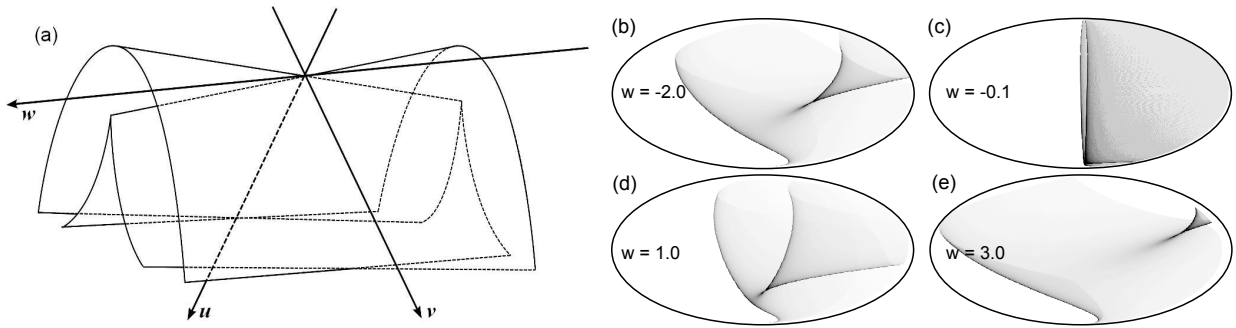}
		\caption{(a) Bifurcation set of the hyperbolic umbilic catastrophe. (b)-(e) Bloch sphere projections showing a hyperbolic umbilic catastrophe unfolding through various values of $w$. \label{unfolding}}
\end{figure*}

To demonstrate this consider the full bifurcation set of the hyperbolic umbilic catastrophe shown in Fig.~\ref{unfolding}(a).  By inputting the known canonical form and using Eq.~\eqref{Blochgerm} we can map an isolated hyperbolic umbilic catastrophe to the Bloch sphere and control its unfolding. This is demonstrated in Figs.~\ref{unfolding}(b)--(e).  This method of spin--engineering by way of using a catastrophe's canonical form could be used to generate any Bloch sphere mapping at a given unfolding.
The hyperbolic umbilic catastrophes that appear on the Bloch sphere are intersections of the Bloch sphere surface with the catastrophe's 3--dimensional bifurcation set. In this example the cross--sections are the $u-v$ plane. As the control parameter $w$ is varied the $u-v$ plane shifts along the $w$ axis, which unfolds the Bloch sphere catastrophe. In general, a hyperbolic umbilic Bloch sphere catastrophe will express itself as cross--sections of the bifurcation set through some arbitrary plane, in which case the unfolding of the catastrophe will depend on all three control parameters $u$, $v$ and $w$. 

\section{Discussion}
Order--parameter catastrophe defects are a generic form of topological defect that are found in fields.
Catastrophes are very general structures that almost always arise in non--trivial mappings to an order parameter manifold. Equations~\eqref{gradmap1} and \eqref{gradmap2} show that the order--parameter coordinates are proportional to the first partial derivatives of a catastrophe's canonical form, which are always simple polynomial functions. This means that both $\phi$ and $\theta$ will also have a local polynomial form for regions that map to a catastrophe. If we consider any sufficiently small local expansion about a point of an arbitrary spin field, the patch only need be diffeomorphic to the canonical form of one of the elementary catastrophes. This criterion is satisfied for the creation of a catastrophe and demonstrates that a mapping of a generic field to its order parameter space will result in caustic tattoos on its manifold. This is a general result that can be found in other types of mappings. For example, catastrophes manifest themselves as {\itshape Argand--plane caustics}, which arise when a complex scalar field is mapped to the Argand--plane~\cite{rothschild, Rothschild14}. In that context, the condition that a point maps to a caustic, $\mbox{det}[J(x,y)]=0$, is equivalent to the optical vorticity vanishing. By making a similar comparison in Eq.~\eqref{hessian} for a complex field it is then apparent that Argand--plane catastrophes also rise from gradient mappings.

Whilst we have used the specific example of 2D spin--1/2 fields and the Bloch sphere, it is important to emphasize that all the concepts presented here are very general. We need only ask what is the mapping of any field to its order parameter space to be able to view the field through the new perspective of order parameter catastrophe defects. This concept is not restricted to 2--dimensions; the 2D surface of the Bloch sphere allows us to view 1D or 2D catastrophes, such as folds or cusps, or 2D cross--sections of higher order catastrophes, such as the elliptic or hyperbolic umbilic catastrophes (Fig.~\ref{zeros}). For mappings to order parameter spaces of higher dimensionality, the full bifurcation sets or higher dimensional cross--sections of catastrophes with higher co--dimension would also become observable.

In summary, order parameter catastrophe defects represent a new form of topological defect in fields. Their stability is attributed to the intrinsic topology of the catastrophes themselves rather than that of the order parameter space, as it is the case for standard topological defects and textures.

\begin{acknowledgments}
The authors thank Dr. Kavan Modi for insightful discussions.  S.A.E gratefully acknowledges funding from an Australian Postgraduate Award. D.M.P and M.J.M acknowledge funding from the Australian Research Council.
\end{acknowledgments}


\begin{thebibliography}{37}%
\makeatletter
\providecommand \@ifxundefined [1]{%
 \@ifx{#1\undefined}
}%
\providecommand \@ifnum [1]{%
 \ifnum #1\expandafter \@firstoftwo
 \else \expandafter \@secondoftwo
 \fi
}%
\providecommand \@ifx [1]{%
 \ifx #1\expandafter \@firstoftwo
 \else \expandafter \@secondoftwo
 \fi
}%
\providecommand \natexlab [1]{#1}%
\providecommand \enquote  [1]{``#1''}%
\providecommand \bibnamefont  [1]{#1}%
\providecommand \bibfnamefont [1]{#1}%
\providecommand \citenamefont [1]{#1}%
\providecommand \href@noop [0]{\@secondoftwo}%
\providecommand \href [0]{\begingroup \@sanitize@url \@href}%
\providecommand \@href[1]{\@@startlink{#1}\@@href}%
\providecommand \@@href[1]{\endgroup#1\@@endlink}%
\providecommand \@sanitize@url [0]{\catcode `\\12\catcode `\$12\catcode
  `\&12\catcode `\#12\catcode `\^12\catcode `\_12\catcode `\%12\relax}%
\providecommand \@@startlink[1]{}%
\providecommand \@@endlink[0]{}%
\providecommand \url  [0]{\begingroup\@sanitize@url \@url }%
\providecommand \@url [1]{\endgroup\@href {#1}{\urlprefix }}%
\providecommand \urlprefix  [0]{URL }%
\providecommand \Eprint [0]{\href }%
\providecommand \doibase [0]{http://dx.doi.org/}%
\providecommand \selectlanguage [0]{\@gobble}%
\providecommand \bibinfo  [0]{\@secondoftwo}%
\providecommand \bibfield  [0]{\@secondoftwo}%
\providecommand \translation [1]{[#1]}%
\providecommand \BibitemOpen [0]{}%
\providecommand \bibitemStop [0]{}%
\providecommand \bibitemNoStop [0]{.\EOS\space}%
\providecommand \EOS [0]{\spacefactor3000\relax}%
\providecommand \BibitemShut  [1]{\csname bibitem#1\endcsname}%
\let\auto@bib@innerbib\@empty
\bibitem [{\citenamefont {Sethna}(2006)}]{sethna}%
  \BibitemOpen
  \bibfield  {author} {\bibinfo {author} {\bibfnamefont {J.}~\bibnamefont
  {Sethna}},\ }\href@noop {} {\emph {\bibinfo {title} {Statistical Mechanics:
  Entropy, Order Parameters and Complexity}}}\ (\bibinfo  {publisher} {Oxford
  University Press},\ \bibinfo {address} {Oxford},\ \bibinfo {year}
  {2006})\BibitemShut {NoStop}%
\bibitem [{\citenamefont {Mermin}(1979)}]{mermin}%
  \BibitemOpen
  \bibfield  {author} {\bibinfo {author} {\bibfnamefont {N.~D.}\ \bibnamefont
  {Mermin}},\ }\href@noop {} {\bibfield  {journal} {\bibinfo  {journal} {Rev.
  Mod. Phys.}\ }\textbf {\bibinfo {volume} {51}},\ \bibinfo {pages} {591}
  (\bibinfo {year} {1979})}\BibitemShut {NoStop}%
\bibitem [{\citenamefont {Golovach}\ \emph {et~al.}(2010)\citenamefont
  {Golovach}, \citenamefont {Borhani},\ and\ \citenamefont {Loss}}]{golovach}%
  \BibitemOpen
  \bibfield  {author} {\bibinfo {author} {\bibfnamefont {V.~N.}\ \bibnamefont
  {Golovach}}, \bibinfo {author} {\bibfnamefont {M.}~\bibnamefont {Borhani}}, \
  and\ \bibinfo {author} {\bibfnamefont {D.}~\bibnamefont {Loss}},\ }\href@noop
  {} {\bibfield  {journal} {\bibinfo  {journal} {Phys. Rev. A}\ }\textbf
  {\bibinfo {volume} {81}},\ \bibinfo {pages} {022315} (\bibinfo {year}
  {2010})}\BibitemShut {NoStop}%
\bibitem [{\citenamefont {Pieranski}\ \emph {et~al.}(2013)\citenamefont
  {Pieranski}, \citenamefont {Yang}, \citenamefont {Burtz}, \citenamefont
  {Camu},\ and\ \citenamefont {Simonetti}}]{liquidcrystal}%
  \BibitemOpen
  \bibfield  {author} {\bibinfo {author} {\bibfnamefont {P.}~\bibnamefont
  {Pieranski}}, \bibinfo {author} {\bibfnamefont {B.}~\bibnamefont {Yang}},
  \bibinfo {author} {\bibfnamefont {L.~J.}\ \bibnamefont {Burtz}}, \bibinfo
  {author} {\bibfnamefont {A.}~\bibnamefont {Camu}}, \ and\ \bibinfo {author}
  {\bibfnamefont {F.}~\bibnamefont {Simonetti}},\ }\href {\doibase
  10.1080/02678292.2012.742581} {\bibfield  {journal} {\bibinfo  {journal}
  {Liq. Cryst.}\ }\textbf {\bibinfo {volume} {40}},\ \bibinfo {pages} {1593}
  (\bibinfo {year} {2013})}\BibitemShut {NoStop}%
\bibitem [{\citenamefont {R$\ddot{\mbox{o}}$szler}\ \emph
  {et~al.}(2006)\citenamefont {R$\ddot{\mbox{o}}$szler}, \citenamefont
  {Bogdanov},\ and\ \citenamefont {Pflederer}}]{pflederer}%
  \BibitemOpen
  \bibfield  {author} {\bibinfo {author} {\bibfnamefont {U.~K.}\ \bibnamefont
  {R$\ddot{\mbox{o}}$szler}}, \bibinfo {author} {\bibfnamefont {A.~N.}\
  \bibnamefont {Bogdanov}}, \ and\ \bibinfo {author} {\bibfnamefont
  {C.}~\bibnamefont {Pflederer}},\ }\href@noop {} {\bibfield  {journal}
  {\bibinfo  {journal} {Nature}\ }\textbf {\bibinfo {volume} {442}},\ \bibinfo
  {pages} {797} (\bibinfo {year} {2006})}\BibitemShut {NoStop}%
\bibitem [{\citenamefont {M{\"u}hlbauer}\ \emph {et~al.}(2009)\citenamefont
  {M{\"u}hlbauer}, \citenamefont {Binz}, \citenamefont {Jonietz}, \citenamefont
  {Pfleiderer}, \citenamefont {Rosch}, \citenamefont {Neubauer}, \citenamefont
  {Georgii},\ and\ \citenamefont {B{\"o}ni}}]{muhlbauer2009skyrmion}%
  \BibitemOpen
  \bibfield  {author} {\bibinfo {author} {\bibfnamefont {S.}~\bibnamefont
  {M{\"u}hlbauer}}, \bibinfo {author} {\bibfnamefont {B.}~\bibnamefont {Binz}},
  \bibinfo {author} {\bibfnamefont {F.}~\bibnamefont {Jonietz}}, \bibinfo
  {author} {\bibfnamefont {C.}~\bibnamefont {Pfleiderer}}, \bibinfo {author}
  {\bibfnamefont {A.}~\bibnamefont {Rosch}}, \bibinfo {author} {\bibfnamefont
  {A.}~\bibnamefont {Neubauer}}, \bibinfo {author} {\bibfnamefont
  {R.}~\bibnamefont {Georgii}}, \ and\ \bibinfo {author} {\bibfnamefont
  {P.}~\bibnamefont {B{\"o}ni}},\ }\href@noop {} {\bibfield  {journal}
  {\bibinfo  {journal} {Science}\ }\textbf {\bibinfo {volume} {323}},\ \bibinfo
  {pages} {915} (\bibinfo {year} {2009})}\BibitemShut {NoStop}%
\bibitem [{\citenamefont {Ezwana}(2010)}]{Ezwana}%
  \BibitemOpen
  \bibfield  {author} {\bibinfo {author} {\bibfnamefont {D.}~\bibnamefont
  {Ezwana}},\ }\href@noop {} {\bibfield  {journal} {\bibinfo  {journal} {Phys.
  Rev. Lett.}\ }\textbf {\bibinfo {volume} {105}},\ \bibinfo {pages} {197202}
  (\bibinfo {year} {2010})}\BibitemShut {NoStop}%
\bibitem [{\citenamefont {Wang}\ \emph {et~al.}(2003)\citenamefont {Wang} \emph
  {et~al.}}]{wang2003epitaxial}%
  \BibitemOpen
  \bibfield  {author} {\bibinfo {author} {\bibfnamefont {J.}~\bibnamefont
  {Wang}} \emph {et~al.},\ }\href@noop {} {\bibfield  {journal} {\bibinfo
  {journal} {Science}\ }\textbf {\bibinfo {volume} {299}},\ \bibinfo {pages}
  {1719} (\bibinfo {year} {2003})}\BibitemShut {NoStop}%
\bibitem [{\citenamefont {Kimura}\ \emph {et~al.}(2003)\citenamefont {Kimura},
  \citenamefont {Kawamoto}, \citenamefont {Yamada}, \citenamefont {Azuma},
  \citenamefont {Takano},\ and\ \citenamefont {Tokura}}]{kimura2003}%
  \BibitemOpen
  \bibfield  {author} {\bibinfo {author} {\bibfnamefont {T.}~\bibnamefont
  {Kimura}}, \bibinfo {author} {\bibfnamefont {S.}~\bibnamefont {Kawamoto}},
  \bibinfo {author} {\bibfnamefont {I.}~\bibnamefont {Yamada}}, \bibinfo
  {author} {\bibfnamefont {M.}~\bibnamefont {Azuma}}, \bibinfo {author}
  {\bibfnamefont {M.}~\bibnamefont {Takano}}, \ and\ \bibinfo {author}
  {\bibfnamefont {Y.}~\bibnamefont {Tokura}},\ }\href@noop {} {\bibfield
  {journal} {\bibinfo  {journal} {Phys. Rev. B}\ }\textbf {\bibinfo {volume}
  {67}},\ \bibinfo {pages} {180401} (\bibinfo {year} {2003})}\BibitemShut
  {NoStop}%
\bibitem [{\citenamefont {Chu}\ \emph {et~al.}(2008)\citenamefont {Chu} \emph
  {et~al.}}]{chu2008electric}%
  \BibitemOpen
  \bibfield  {author} {\bibinfo {author} {\bibfnamefont {Y.~H.}\ \bibnamefont
  {Chu}} \emph {et~al.},\ }\href@noop {} {\bibfield  {journal} {\bibinfo
  {journal} {Nature Mat.}\ }\textbf {\bibinfo {volume} {7}},\ \bibinfo {pages}
  {478} (\bibinfo {year} {2008})}\BibitemShut {NoStop}%
\bibitem [{\citenamefont {Gull}\ \emph {et~al.}(2013)\citenamefont {Gull},
  \citenamefont {Parcollet},\ and\ \citenamefont
  {Millis}}]{PhysRevLett.110.216405}%
  \BibitemOpen
  \bibfield  {author} {\bibinfo {author} {\bibfnamefont {E.}~\bibnamefont
  {Gull}}, \bibinfo {author} {\bibfnamefont {O.}~\bibnamefont {Parcollet}}, \
  and\ \bibinfo {author} {\bibfnamefont {A.~J.}\ \bibnamefont {Millis}},\
  }\href@noop {} {\bibfield  {journal} {\bibinfo  {journal} {Phys. Rev. Lett.}\
  }\textbf {\bibinfo {volume} {110}},\ \bibinfo {pages} {216405} (\bibinfo
  {year} {2013})}\BibitemShut {NoStop}%
\bibitem [{\citenamefont {Seibold}\ \emph {et~al.}(2012)\citenamefont
  {Seibold}, \citenamefont {Benfatto}, \citenamefont {Castellani},\ and\
  \citenamefont {Lorenzana}}]{PhysRevLett.108.207004}%
  \BibitemOpen
  \bibfield  {author} {\bibinfo {author} {\bibfnamefont {G.}~\bibnamefont
  {Seibold}}, \bibinfo {author} {\bibfnamefont {L.}~\bibnamefont {Benfatto}},
  \bibinfo {author} {\bibfnamefont {C.}~\bibnamefont {Castellani}}, \ and\
  \bibinfo {author} {\bibfnamefont {J.}~\bibnamefont {Lorenzana}},\ }\href@noop
  {} {\bibfield  {journal} {\bibinfo  {journal} {Phys. Rev. Lett.}\ }\textbf
  {\bibinfo {volume} {108}},\ \bibinfo {pages} {207004} (\bibinfo {year}
  {2012})}\BibitemShut {NoStop}%
\bibitem [{\citenamefont {Fairbairn}\ and\ \citenamefont
  {Hogan}(2013)}]{particlecomo}%
  \BibitemOpen
  \bibfield  {author} {\bibinfo {author} {\bibfnamefont {M.}~\bibnamefont
  {Fairbairn}}\ and\ \bibinfo {author} {\bibfnamefont {R.}~\bibnamefont
  {Hogan}},\ }\href@noop {} {\bibfield  {journal} {\bibinfo  {journal} {J. High
  Energy Phys.}\ }\textbf {\bibinfo {volume} {2013}},\ \bibinfo {pages} {1}
  (\bibinfo {year} {2013})}\BibitemShut {NoStop}%
\bibitem [{\citenamefont {Ruben}\ \emph {et~al.}(2010)\citenamefont {Ruben},
  \citenamefont {Morgan},\ and\ \citenamefont {Paganin}}]{ruben2010}%
  \BibitemOpen
  \bibfield  {author} {\bibinfo {author} {\bibfnamefont {G.}~\bibnamefont
  {Ruben}}, \bibinfo {author} {\bibfnamefont {M.~J.}\ \bibnamefont {Morgan}}, \
  and\ \bibinfo {author} {\bibfnamefont {D.~M.}\ \bibnamefont {Paganin}},\
  }\href {\doibase 10.1103/PhysRevLett.105.220402} {\bibfield  {journal}
  {\bibinfo  {journal} {Phys. Rev. Lett.}\ }\textbf {\bibinfo {volume} {105}},\
  \bibinfo {pages} {220402} (\bibinfo {year} {2010})}\BibitemShut {NoStop}%
\bibitem [{\citenamefont {Arnol'd}(1975)}]{arnold1975}%
  \BibitemOpen
  \bibfield  {author} {\bibinfo {author} {\bibfnamefont {V.~I.}\ \bibnamefont
  {Arnol'd}},\ }\href@noop {} {\bibfield  {journal} {\bibinfo  {journal} {Russ.
  Math. Surv+}\ }\textbf {\bibinfo {volume} {30}},\ \bibinfo {pages} {1}
  (\bibinfo {year} {1975})}\BibitemShut {NoStop}%
\bibitem [{\citenamefont {Saunders}(1980)}]{saunders}%
  \BibitemOpen
  \bibfield  {author} {\bibinfo {author} {\bibfnamefont {P.~T.}\ \bibnamefont
  {Saunders}},\ }\href@noop {} {\emph {\bibinfo {title} {An Introduction To
  Catastrophe Theory}}}\ (\bibinfo  {publisher} {Cambridge University Press},\
  \bibinfo {address} {Cambridge},\ \bibinfo {year} {1980})\BibitemShut
  {NoStop}%
\bibitem [{\citenamefont {Poston}\ and\ \citenamefont
  {Steward}(1996)}]{poston}%
  \BibitemOpen
  \bibfield  {author} {\bibinfo {author} {\bibfnamefont {T.}~\bibnamefont
  {Poston}}\ and\ \bibinfo {author} {\bibfnamefont {I.}~\bibnamefont
  {Steward}},\ }\href@noop {} {\emph {\bibinfo {title} {Catastrophe Theory and
  Its Applications}}}\ (\bibinfo  {publisher} {Dover Publications Inc.},\
  \bibinfo {address} {New York},\ \bibinfo {year} {1996})\BibitemShut {NoStop}%
\bibitem [{\citenamefont {Thom}(1983)}]{thom1983}%
  \BibitemOpen
  \bibfield  {author} {\bibinfo {author} {\bibfnamefont {R.}~\bibnamefont
  {Thom}},\ }\href@noop {} {\emph {\bibinfo {title} {Mathematical Models of
  Morphogenesis}}}\ (\bibinfo  {publisher} {Wiley},\ \bibinfo {address} {New
  York},\ \bibinfo {year} {1983})\BibitemShut {NoStop}%
\bibitem [{\citenamefont {Hannay}(1982)}]{hannay1982}%
  \BibitemOpen
  \bibfield  {author} {\bibinfo {author} {\bibfnamefont {J.}~\bibnamefont
  {Hannay}},\ }\href {\doibase 10.1080/713820808} {\bibfield  {journal}
  {\bibinfo  {journal} {Opt. Acta}\ }\textbf {\bibinfo {volume} {29}},\
  \bibinfo {pages} {1631} (\bibinfo {year} {1982})}\BibitemShut {NoStop}%
\bibitem [{\citenamefont {Berry}(1976)}]{berry1}%
  \BibitemOpen
  \bibfield  {author} {\bibinfo {author} {\bibfnamefont {M.~V.}\ \bibnamefont
  {Berry}},\ }\href@noop {} {\bibfield  {journal} {\bibinfo  {journal} {Adv.
  Phys.}\ }\textbf {\bibinfo {volume} {25}},\ \bibinfo {pages} {1} (\bibinfo
  {year} {1976})}\BibitemShut {NoStop}%
\bibitem [{\citenamefont {Berry}\ and\ \citenamefont {Upstill}(1980)}]{berry2}%
  \BibitemOpen
  \bibfield  {author} {\bibinfo {author} {\bibfnamefont {M.~V.}\ \bibnamefont
  {Berry}}\ and\ \bibinfo {author} {\bibfnamefont {C.}~\bibnamefont
  {Upstill}},\ }\href@noop {} {\bibfield  {journal} {\bibinfo  {journal} {Prog.
  Opt.}\ }\textbf {\bibinfo {volume} {18}},\ \bibinfo {pages} {259} (\bibinfo
  {year} {1980})}\BibitemShut {NoStop}%
\bibitem [{\citenamefont {Nye}(1978)}]{nye1978}%
  \BibitemOpen
  \bibfield  {author} {\bibinfo {author} {\bibfnamefont {J.}~\bibnamefont
  {Nye}},\ }\href@noop {} {\bibfield  {journal} {\bibinfo  {journal} {Proc. R.
  Soc. A}\ }\textbf {\bibinfo {volume} {361}},\ \bibinfo {pages} {21} (\bibinfo
  {year} {1978})}\BibitemShut {NoStop}%
\bibitem [{\citenamefont {Nye}(2003{\natexlab{a}})}]{nye2003a}%
  \BibitemOpen
  \bibfield  {author} {\bibinfo {author} {\bibfnamefont {J.~F.}\ \bibnamefont
  {Nye}},\ }\href@noop {} {\bibfield  {journal} {\bibinfo  {journal} {J. Opt.
  A-Pure Appl. Op.}\ }\textbf {\bibinfo {volume} {5}},\ \bibinfo {pages} {495}
  (\bibinfo {year} {2003}{\natexlab{a}})}\BibitemShut {NoStop}%
\bibitem [{\citenamefont {Nye}(2003{\natexlab{b}})}]{nye2003b}%
  \BibitemOpen
  \bibfield  {author} {\bibinfo {author} {\bibfnamefont {J.~F.}\ \bibnamefont
  {Nye}},\ }\href@noop {} {\bibfield  {journal} {\bibinfo  {journal} {J. Opt.
  A-Pure Appl. Op.}\ }\textbf {\bibinfo {volume} {5}},\ \bibinfo {pages} {503}
  (\bibinfo {year} {2003}{\natexlab{b}})}\BibitemShut {NoStop}%
\bibitem [{\citenamefont {Nye}(2006)}]{nye2006}%
  \BibitemOpen
  \bibfield  {author} {\bibinfo {author} {\bibfnamefont {J.~F.}\ \bibnamefont
  {Nye}},\ }\href@noop {} {\bibfield  {journal} {\bibinfo  {journal} {J. Opt.
  A-Pure Appl. Op.}\ }\textbf {\bibinfo {volume} {8}},\ \bibinfo {pages} {304}
  (\bibinfo {year} {2006})}\BibitemShut {NoStop}%
\bibitem [{\citenamefont {Petersen}\ \emph {et~al.}(2013)\citenamefont
  {Petersen}, \citenamefont {Weyland}, \citenamefont {Paganin}, \citenamefont
  {Simula}, \citenamefont {Eastwood},\ and\ \citenamefont
  {Morgan}}]{petersen2013}%
  \BibitemOpen
  \bibfield  {author} {\bibinfo {author} {\bibfnamefont {T.~C.}\ \bibnamefont
  {Petersen}}, \bibinfo {author} {\bibfnamefont {M.}~\bibnamefont {Weyland}},
  \bibinfo {author} {\bibfnamefont {D.~M.}\ \bibnamefont {Paganin}}, \bibinfo
  {author} {\bibfnamefont {T.~P.}\ \bibnamefont {Simula}}, \bibinfo {author}
  {\bibfnamefont {S.~A.}\ \bibnamefont {Eastwood}}, \ and\ \bibinfo {author}
  {\bibfnamefont {M.~J.}\ \bibnamefont {Morgan}},\ }\href@noop {} {\bibfield
  {journal} {\bibinfo  {journal} {Phys. Rev. Lett.}\ }\textbf {\bibinfo
  {volume} {110}},\ \bibinfo {pages} {033901} (\bibinfo {year}
  {2013})}\BibitemShut {NoStop}%
\bibitem [{\citenamefont {Simula}\ \emph {et~al.}(2013)\citenamefont {Simula},
  \citenamefont {Petersen},\ and\ \citenamefont {Paganin}}]{simula2013}%
  \BibitemOpen
  \bibfield  {author} {\bibinfo {author} {\bibfnamefont {T.~P.}\ \bibnamefont
  {Simula}}, \bibinfo {author} {\bibfnamefont {T.~C.}\ \bibnamefont
  {Petersen}}, \ and\ \bibinfo {author} {\bibfnamefont {D.~M.}\ \bibnamefont
  {Paganin}},\ }\href@noop {} {\bibfield  {journal} {\bibinfo  {journal} {Phys.
  Rev. A}\ }\textbf {\bibinfo {volume} {88}},\ \bibinfo {pages} {043626}
  (\bibinfo {year} {2013})}\BibitemShut {NoStop}%
\bibitem [{\citenamefont {Thom}(1977)}]{thom1977}%
  \BibitemOpen
  \bibfield  {author} {\bibinfo {author} {\bibfnamefont {R.}~\bibnamefont
  {Thom}},\ }\href {\doibase 10.1137/1019036} {\bibfield  {journal} {\bibinfo
  {journal} {SIAM Rev.}\ }\textbf {\bibinfo {volume} {19}},\ \bibinfo {pages}
  {189} (\bibinfo {year} {1977})}\BibitemShut {NoStop}%
\bibitem [{\citenamefont {Maeda}\ \emph {et~al.}(1994)\citenamefont {Maeda},
  \citenamefont {Tachizawa}, \citenamefont {Torii},\ and\ \citenamefont
  {Maki}}]{maeda1994}%
  \BibitemOpen
  \bibfield  {author} {\bibinfo {author} {\bibfnamefont {K.}~\bibnamefont
  {Maeda}}, \bibinfo {author} {\bibfnamefont {T.}~\bibnamefont {Tachizawa}},
  \bibinfo {author} {\bibfnamefont {T.}~\bibnamefont {Torii}}, \ and\ \bibinfo
  {author} {\bibfnamefont {T.}~\bibnamefont {Maki}},\ }\href@noop {} {\bibfield
   {journal} {\bibinfo  {journal} {Phys. Rev. Lett.}\ }\textbf {\bibinfo
  {volume} {72}},\ \bibinfo {pages} {450} (\bibinfo {year} {1994})}\BibitemShut
  {NoStop}%
\bibitem [{\citenamefont {Martinis}\ \emph {et~al.}(2003)\citenamefont
  {Martinis}, \citenamefont {Nam}, \citenamefont {Aumentado}, \citenamefont
  {Lang},\ and\ \citenamefont {Urbina}}]{PhysRevB.67.094510}%
  \BibitemOpen
  \bibfield  {author} {\bibinfo {author} {\bibfnamefont {J.~M.}\ \bibnamefont
  {Martinis}}, \bibinfo {author} {\bibfnamefont {S.}~\bibnamefont {Nam}},
  \bibinfo {author} {\bibfnamefont {J.}~\bibnamefont {Aumentado}}, \bibinfo
  {author} {\bibfnamefont {K.~M.}\ \bibnamefont {Lang}}, \ and\ \bibinfo
  {author} {\bibfnamefont {C.}~\bibnamefont {Urbina}},\ }\href@noop {}
  {\bibfield  {journal} {\bibinfo  {journal} {Phys. Rev. B}\ }\textbf {\bibinfo
  {volume} {67}},\ \bibinfo {pages} {094510} (\bibinfo {year}
  {2003})}\BibitemShut {NoStop}%
\bibitem [{\citenamefont {Xiang}\ \emph {et~al.}(2005)\citenamefont {Xiang},
  \citenamefont {Li}, \citenamefont {Yu},\ and\ \citenamefont
  {Guo}}]{PhysRevA.72.012315}%
  \BibitemOpen
  \bibfield  {author} {\bibinfo {author} {\bibfnamefont {G.-Y.}\ \bibnamefont
  {Xiang}}, \bibinfo {author} {\bibfnamefont {J.}~\bibnamefont {Li}}, \bibinfo
  {author} {\bibfnamefont {B.}~\bibnamefont {Yu}}, \ and\ \bibinfo {author}
  {\bibfnamefont {G.-C.}\ \bibnamefont {Guo}},\ }\href@noop {} {\bibfield
  {journal} {\bibinfo  {journal} {Phys. Rev. A}\ }\textbf {\bibinfo {volume}
  {72}},\ \bibinfo {pages} {012315} (\bibinfo {year} {2005})}\BibitemShut
  {NoStop}%
\bibitem [{\citenamefont {Nielsen}\ and\ \citenamefont
  {Chuang}(2010)}]{nielsen2010quantum}%
  \BibitemOpen
  \bibfield  {author} {\bibinfo {author} {\bibfnamefont {M.~A.}\ \bibnamefont
  {Nielsen}}\ and\ \bibinfo {author} {\bibfnamefont {I.~L.}\ \bibnamefont
  {Chuang}},\ }\href@noop {} {\emph {\bibinfo {title} {Quantum Computation and
  Quantum Information}}}\ (\bibinfo  {publisher} {Cambridge University Press},\
  \bibinfo {address} {Cambridge},\ \bibinfo {year} {2010})\BibitemShut
  {NoStop}%
\bibitem [{\citenamefont {Zhang}\ \emph {et~al.}(2009)\citenamefont {Zhang},
  \citenamefont {Ran},\ and\ \citenamefont {Vishwanath}}]{PhysRevB.79.245331}%
  \BibitemOpen
  \bibfield  {author} {\bibinfo {author} {\bibfnamefont {Y.}~\bibnamefont
  {Zhang}}, \bibinfo {author} {\bibfnamefont {Y.}~\bibnamefont {Ran}}, \ and\
  \bibinfo {author} {\bibfnamefont {A.}~\bibnamefont {Vishwanath}},\
  }\href@noop {} {\bibfield  {journal} {\bibinfo  {journal} {Phys. Rev. B}\
  }\textbf {\bibinfo {volume} {79}},\ \bibinfo {pages} {245331} (\bibinfo
  {year} {2009})}\BibitemShut {NoStop}%
\bibitem [{\citenamefont {Kawaguchi}\ \emph {et~al.}(2008)\citenamefont
  {Kawaguchi}, \citenamefont {Nitta},\ and\ \citenamefont
  {Ueda}}]{PhysRevLett.100.180403}%
  \BibitemOpen
  \bibfield  {author} {\bibinfo {author} {\bibfnamefont {Y.}~\bibnamefont
  {Kawaguchi}}, \bibinfo {author} {\bibfnamefont {M.}~\bibnamefont {Nitta}}, \
  and\ \bibinfo {author} {\bibfnamefont {M.}~\bibnamefont {Ueda}},\ }\href@noop
  {} {\bibfield  {journal} {\bibinfo  {journal} {Phys. Rev. Lett.}\ }\textbf
  {\bibinfo {volume} {100}},\ \bibinfo {pages} {180403} (\bibinfo {year}
  {2008})}\BibitemShut {NoStop}%
\bibitem [{\citenamefont {M{\"a}kel{\"a}}\ \emph {et~al.}(2003)\citenamefont
  {M{\"a}kel{\"a}}, \citenamefont {Zhang},\ and\ \citenamefont
  {Suominen}}]{makela2003}%
  \BibitemOpen
  \bibfield  {author} {\bibinfo {author} {\bibfnamefont {H.}~\bibnamefont
  {M{\"a}kel{\"a}}}, \bibinfo {author} {\bibfnamefont {Y.}~\bibnamefont
  {Zhang}}, \ and\ \bibinfo {author} {\bibfnamefont {K.-A.}\ \bibnamefont
  {Suominen}},\ }\href@noop {} {\bibfield  {journal} {\bibinfo  {journal} {J.
  Phys. A}\ }\textbf {\bibinfo {volume} {36}},\ \bibinfo {pages} {8555}
  (\bibinfo {year} {2003})}\BibitemShut {NoStop}%
\bibitem [{\citenamefont {Rothschild}\ \emph {et~al.}(2012)\citenamefont
  {Rothschild}, \citenamefont {Kitchen}, \citenamefont {Faulkner},\ and\
  \citenamefont {Paganin}}]{rothschild}%
  \BibitemOpen
  \bibfield  {author} {\bibinfo {author} {\bibfnamefont {F.}~\bibnamefont
  {Rothschild}}, \bibinfo {author} {\bibfnamefont {M.~J.}\ \bibnamefont
  {Kitchen}}, \bibinfo {author} {\bibfnamefont {H.~M.~L.}\ \bibnamefont
  {Faulkner}}, \ and\ \bibinfo {author} {\bibfnamefont {D.~M.}\ \bibnamefont
  {Paganin}},\ }\href@noop {} {\bibfield  {journal} {\bibinfo  {journal}
  {Optics Commun.}\ }\textbf {\bibinfo {volume} {285}},\ \bibinfo {pages}
  {4141} (\bibinfo {year} {2012})}\BibitemShut {NoStop}%
\bibitem [{\citenamefont {Rothschild}\ \emph {et~al.}(2014)\citenamefont
  {Rothschild}, \citenamefont {Bishop}, \citenamefont {Kitchen},\ and\
  \citenamefont {Paganin}}]{Rothschild14}%
  \BibitemOpen
  \bibfield  {author} {\bibinfo {author} {\bibfnamefont {F.}~\bibnamefont
  {Rothschild}}, \bibinfo {author} {\bibfnamefont {A.~I.}\ \bibnamefont
  {Bishop}}, \bibinfo {author} {\bibfnamefont {M.~J.}\ \bibnamefont {Kitchen}},
  \ and\ \bibinfo {author} {\bibfnamefont {D.~M.}\ \bibnamefont {Paganin}},\
  }\href@noop {} {\bibfield  {journal} {\bibinfo  {journal} {Opt. Express}\
  }\textbf {\bibinfo {volume} {22}},\ \bibinfo {pages} {6495} (\bibinfo {year}
  {2014})}\BibitemShut {NoStop}%
\end{thebibliography}

%

\end{document}